\documentclass[
pra,
reprint,
a4paper,
longbibliography,
]{revtex4-1}
\usepackage[utf8]{inputenc}
\usepackage[T1]{fontenc}

\usepackage{amsfonts}
\usepackage{amsmath}
\usepackage{amsthm}
\usepackage{braket}
\usepackage{graphicx}
\usepackage{hyperref}

\hypersetup{colorlinks=true}
\hypersetup{linkcolor=blue}
\hypersetup{citecolor=magenta}
\hypersetup{urlcolor=blue}

\DeclareMathOperator{\tr}{tr}
\theoremstyle{definition}
\newtheorem{theorem}{Theorem}
\newtheorem{corollary}[theorem]{Consequence}
\newcommand{\omitted}{{\rule{1ex}{.4pt}}}
\newcommand{\tast}{{${}^\ast$}}
\providecommand{\naturals}{{\mathbb N}}

\begin{document}
\title{%
Quantum correlations are stronger than all nonsignaling correlations produced 
 by $n$-outcome measurements
}

\author{Matthias~Kleinmann}
\email{matthias\_kleinmann001@ehu.eus}
\affiliation{%
Department of Theoretical Physics,
University of the Basque Country UPV/EHU,
P.O.\ Box 644,
E-48080 Bilbao,
Spain}

\author{Adán~Cabello}
\email{adan@us.es}
\affiliation{%
Departamento de Física Aplicada II,
Universidad de Sevilla,
E-41012 Sevilla,
Spain}

\begin{abstract}
We show that, for any $n$, there are $m$-outcome quantum correlations, with 
 $m>n$, which are stronger than any nonsignaling correlation produced from 
 selecting among $n$-outcome measurements.
As a consequence, for any $n$, there are $m$-outcome quantum measurements that 
 cannot be constructed by selecting locally from the set of $n$-outcome 
 measurements.
This is a property of the set of measurements in quantum theory that is not 
 mandatory for general probabilistic theories.
We also show that this prediction can be tested through high-precision 
 Bell-type experiments and identify past experiments providing evidence that 
 some of these strong correlations exist in nature.
Finally, we provide a modified version of quantum theory restricted to having 
 at most $n$-outcome quantum measurements.
\end{abstract}

\maketitle

\textit{Introduction.}---%
The violation of Bell inequalities \cite{Freedman:1972PRL, Aspect:1982PRL, 
 Weihs:1998PRL, Hensen:2015NAT, Giustina:2015PRL, Shalm:2015PRL} does not only 
 show the impossibility of local realism \cite{Bell:1964PHY}, but also 
 demonstrates
(i) the existence of entangled states, i.e., states which cannot be produced by 
 choosing among states produced locally, and
(ii) the existence of incompatible measurements, i.e., measurements whose 
 outcomes cannot be obtained from a single joint measurement.
Remarkably, this holds not only assuming quantum theory (QT) but also holds for 
 the much broader set of general probabilistic theories (GPTs) 
 \cite{Ludwig:1987, Mittelstaedt:1998, Chiribella:2010PRA, Acin:2010PRL}.
GPTs include classical probability theory and QT, and also theories admitting 
 supraquantum nonsignaling correlations, such as, e.g., Popescu-Rohrlich boxes 
 \cite{Popescu:1994FPH}.

Svetlichny pointed out that (i) can be refined and that for any number of 
 parties $n$, there are correlations predicted by QT that cannot be explained 
 by any GPT in which all states are produced by choosing among $(n-1)$-partite 
 entangled states \cite{Svetlichny:1987PRD, Collins:2002PRL:2, 
 Seevinck:2002PRL}.
Hence, the violation of $n$-partite Svetlichny inequalities 
 \cite{Lavoie:2009NJP, Erven:2014NPHO, Hamel:2014NPHO, Barreiro:2013NPHY} 
 demonstrates the existence of genuinely $n$-partite entangled states, and 
 therefore puts strong constraints on which GPTs are suitable to describe 
 nature.

\begin{figure}
\includegraphics[width=\linewidth]{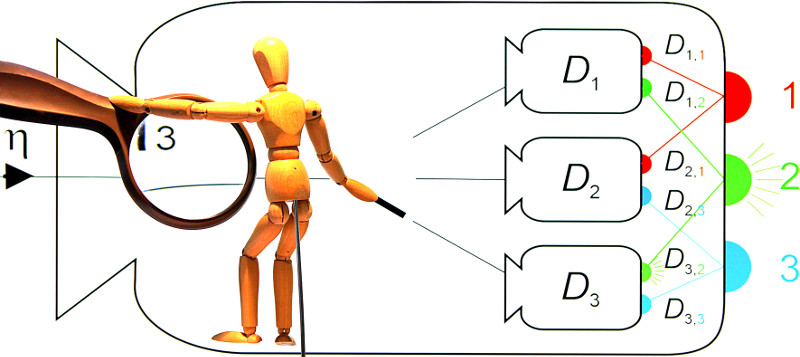}
\caption{%
\label{fig:alice}%
Illustration of a three-outcome measurement which can be explained as selecting 
 one from three two-outcome measurements.
From the outside, the measurement apparatus (represented by the outer box) has 
 three outcomes (represented by three lights of different colors).
The state of a physical system tested by the apparatus is described by 
 $\eta_\alpha$, where $\alpha=1,2,3$ is a variable that is hidden to the 
 experimenter but can be read off by the measurement apparatus (illustrated by 
 a robot inside the box using a magnifying glass), without disturbing the state 
 of the system.
From the inside, the measurement apparatus works as follows: based on the value 
 of $\alpha$ (here: $\alpha=3$) a corresponding two-outcome measurement 
 $D_\alpha$ is selected (as the robot does by operating the switch selecting 
 the measurement $D_3$).}
\end{figure}

Here we address the problem of whether there is a sensible way to go beyond 
 (ii) and, assuming that QT is correct, constrain more rigidly the structure of 
 the set of measurements in any GPT describing nature.
Our main result is the proof that, according to QT, nature does produce 
 correlations which cannot be generated by shared randomness (e.g., by means of 
 local hidden variables) and nonsignaling correlations for which the number of 
 outcomes is limited to $n$.
In this sense, we show that quantum correlations are not $n$-chotomic, for any 
 $n=2,3,\dotsc$.
This implies that, the same way Bell inequality experiments exclude all local 
 realistic theories, QT predicts that certain experiments can exclude all GPTs 
 in which measurements are locally selected from $n$-outcome measurements.
A possible selection mechanism, in which all measurements are produced from 
 two-outcome measurements with the help of hidden variables, is illustrated in 
 Fig.~\ref{fig:alice}.

However, according to our analysis, such experiments require visibilities 
 beyond what is currently feasible.
This motivates us to consider a particular subclass of GPTs: those in which 
 measurements are locally selected from $n$-outcome \emph{quantum} 
 measurements.
We identify past experiments which, for $n=2$ and $n=3$ and under some 
 assumptions, may be taken as experimental falsifications of this subclass of 
 GPTs.
Finally, we take the possibility seriously that QT does not account for 
 correlations in nature and provide a modified version of QT restricted to 
 having at most $n$-outcome quantum measurements.
This theory shows that nonsignaling correlations for which the number of 
 outcomes is limited to $n$ constitute an alternative that should be 
 experimentally tested.

\textit{Quantum correlations are not $n$-chotomic.}---%
For $\text{$m>n$}$, the set of $m$-outcome measurements in QT is strictly 
 larger than the convex hull of the $n$-outcome measurements \cite{Busch:1995}.
Hence, there are, e.g., three-outcome quantum measurements which cannot be 
 implemented by choosing one from a set of two-outcome quantum measurements.
Here we present a result which goes beyond this observation.
We demonstrate that if QT is correct, then any GPT describing nature needs to 
 share this property.
For this, we prove the yet more general result that any GPT not having this 
 property cannot reproduce the correlations predicted by QT.
This result only depends on properties of correlations and does not rely on how 
 the preparation and measurement devices work.
Therefore, it enables us to exclude all those GPTs in a device-independent way.

Suppose that two parties can perform several measurements on a bipartite system 
 and that each party can independently choose among the measurement settings.
For a fixed measurement setting $\mu$ on the first party and $\nu$ on the 
 second party, we write $P_{\mu,\nu}(k,\ell)$ for the probability to obtain the 
 corresponding outcomes $k$ and $\ell$.
A set of such correlations is nonsignaling, if $\sum_\ell 
 P_{\mu,\nu}(k,\ell)\equiv P_{\mu,\omitted}(k)$ is independent of $\nu$ and 
 $\sum_k P_{\mu,\nu}(k,\ell)\equiv P_{\omitted,\nu}(\ell)$ is independent of 
 $\mu$.
We are now interested in the case where the number of measurement outcomes is 
 limited to $n$, i.e., the measurements are $n$-chotomic.
An $n$-chotomic local measurement obeys $P_{\mu,\omitted}(k)= 0$ for all $k$, 
 except for a subset of size $n$, or, similarly, $P_{\omitted,\nu}(\ell)= 0$ 
 for all $\ell$, except for a subset of size $n$.
The set of nonsignaling $n$-chotomic correlations $\mathcal P_n$ is then the 
 convex hull of the set of nonsignaling correlations where all measurements 
 are, at most, $n$-chotomic.

We address the question of whether the set of quantum correlations contains 
 correlations that are not in $\mathcal P_n$ by considering the combinations of 
 correlations in the Collins–Gisin–Linden–Massar–Popescu inequalities 
 \cite{Collins:2002PRL} in the formulation of Zohren and Gill 
 \cite{Zohren:2008PRL}, namely,
\begin{equation}\begin{split}
 I'(\boldsymbol P)= P_{2,2}(k<\ell)+ P_{1,2}(k>\ell) + P_{1,1}(k<\ell) \\
 + P_{2,1}(k\ge \ell),
\end{split}\end{equation}
 where $P_{2,2}(k<\ell)= \sum_{k<\ell} P_{2,2}(k,\ell)$, and similarly for the 
 other terms.
$I'$ can be evaluated for any set of bipartite correlations $\boldsymbol P$ 
 which features at least two measurement settings per party.
We can now state our main result.
\begin{theorem}
\label{thm}
For any $n$, there is an $m>n$ and quantum correlations $\boldsymbol Q\in 
 \mathcal P_m$, such that $I'(\boldsymbol Q) < \inf[I'(\mathcal P_n)]$.
\end{theorem}
\begin{proof}
It has been shown \cite{Zohren:2010EPL} that for any $\varepsilon> 0$, there 
 exists an $m$ and some quantum correlations $\boldsymbol Q\in \mathcal P_m$ 
 such that $I'(\boldsymbol Q)< \varepsilon$.
In Appendix~\ref{app:thm} we prove that $q_n\equiv \inf[I'(\mathcal P_n)]>0$ 
 for any $n$.
Therefore, by choosing $\varepsilon= q_n/2$, the assertion follows.
\end{proof}

This proves that, for any $n$, there are quantum correlations which are not 
 nonsignaling $n$-chotomic.
For example, the hypothetical Popescu–Rohrlich box \cite{Popescu:1994FPH} is a 
 GPT predicting correlations that are impossible according to QT.
However, this GPT only contains dichotomic measurements.
Hence, Theorem~\ref{thm} reveals that QT contains correlations that are 
 impossible to achieve for a Popescu–Rohrlich box.
\begin{corollary}
QT contains correlations that cannot be explained by dichotomic GPTs, even if 
 we admit supraquantum correlations, such as Popescu–Rohrlich boxes.
\end{corollary}

\textit{Experiments.}---%
Theorem~\ref{thm} gives rise to the question: Is it feasible to experimentally 
 demonstrate the existence of correlations which cannot be explained by 
 $n$-chotomic GPTs with current quantum technology?
As shown in Theorem~\ref{thm}, in principle, we could use experiments aiming to 
 violate $I'$ for this purpose.
However, in practice, this approach is rather unfeasible since, even for 
 excluding dichotomic GPTs, we would need to observe a value of $I'$ below 
 $\frac12$, something that requires quantum measurements with at least ten 
 outcomes \cite{Zohren:2008PRL}.
Further investigation is therefore needed in order to identify inequalities 
 with more modest experimental demands.

As a first step in this direction, we explore whether it is possible to 
 experimentally exclude GPTs in which measurements are produced by selecting 
 from, at most, $n$-outcome \emph{quantum} measurements.
These GPTs constitute interesting variants of QT in which the sets of 
 measurements are arguably simpler than the one of QT, as we discuss below.
In addition, unlike most alternatives to QT investigated in the past (e.g., 
 local realistic theories), they share most of the predictions of QT, including 
 the violation of Bell inequalities.

\begin{table}
\begin{tabular}{|l|cccc|}\hline
 & VB \cite{Vertesi:2010PRA}
 & $I_3$ \cite{Collins:2002PRL, Acin:2002PRA}
 & $I_4$ \cite{Collins:2002PRL}
 & AN \cite{Cabello:2001PRL}
\\\hline
{2}-outcome
 & 21.068\tast
 & 0.20711\phantom\tast 
 & 0.20711\phantom\tast 
 & 8.1962\phantom\tast 
\\
{3}-outcome
 & ---
 & ---
 & 0.30495\phantom\tast 
 & 8.1962\phantom\tast 
\\
Quantum
 & 21.090\tast
 & 0.30495\tast 
 & 0.36476\tast
 & 9.0000\tast
\\
{2}-visibility
 & 99.97\%
 & 90\%
 & 86\%
 & 92\%
\\
{3}-visibility
 & ---
 & ---
 & 95\%
 & 92\%
\\
Experiment
 & none
 & Ref.~[\onlinecite{Vaziri:2002PRL}]
 & Ref.~[\onlinecite{Ikuta:2016PRA}]
 & Ref.~[\onlinecite{Yang:2005PRL}]
\\
{2}-violation &
 &
 5.5$\sigma$ &
 16$\sigma$ &
 70$\sigma$ \\
{3}-violation
 & ---
 & ---
 & 4.3$\sigma$
 & 70$\sigma$
\\\hline
\end{tabular}
\caption{%
\label{tab:res}%
Upper bounds on correlations, required visibility, and experimental results.
Values with an asterisk\tast\ have been established in prior work.
VB stands for the combination of correlations in the Vértesi–Bene 
 inequality \cite{Vertesi:2010PRA}, $I_n$ for $1-I'$ with possible outcomes 
 $k,\ell=1,2,\dotsc,n$, and AN for the correlations in an all-versus-nothing 
 inequality \cite{Cabello:2001PRL}.
The rows ``{2}-outcome,'' ``{3}-outcome,'' and ``Quantum'' list upper bounds 
 when quantum measurements have only two, only three or an unrestricted number 
 of outcomes, respectively.
In the rows ``{2}-visibility'' and ``{3}-visibility'' the required visibility 
 (absence of white noise, i.e., minimal $p$ if the prepared state is a mixture 
 of the target state and a completely depolarized state, 
 $\varrho_\mathrm{prepared}= p\varrho_\mathrm{target}+ 
 (1-p)\varrho_\mathrm{depolarized}$) is shown, where the former is for 
 violating the two-outcome bound and the latter for violating the three-outcome 
 bound.
In the last rows, observed experimental violations of the two-outcome and 
 three-outcome bounds are shown, in terms of multiples of statistical standard 
 deviations.}
\end{table}

For this purpose, we compute the upper bounds on $1-I'$ for GPTs for dichotomic 
 and trichotomic quantum measurements when the outcomes $k,l$ take values 
 $1,2,3$ ($I_3)$ or values $1,2,3,4$ ($I_4$).
We observe that although violating the resulting inequalities is experimentally 
 demanding, there is already experimental evidence \cite{Vaziri:2002PRL, 
 Dada:2011NPHY, Ikuta:2016PRA} supporting that there are measurements which 
 cannot be explained choosing from quantum dichotomic or quantum trichotomic 
 measurements.
Interestingly, when we compute the upper bounds for the bipartite 
 all-versus-nothing Bell inequality with three four-outcome measurements 
 \cite{Cabello:2001PRL}, we observe that the results of a previous experiment, 
 show a clear violation of the quantum trichotomic bound \cite{Yang:2005PRL}.
This suggests that this inequality can be a powerful tool to provide conclusive 
 evidence of the existence of genuinely nontrichotomic quantum measurements.
We also compute the upper bounds of an inequality due to Vértesi and Bene 
 \cite{Vertesi:2010PRA} which, so far, has not been tested experimentally.
However, it is \emph{a priori} interesting for our considerations, since it can 
 be violated by a two-qubit system.
Unfortunately, we find that the visibility required to falsify dichotomic 
 quantum measurements using the Vértesi–Bene inequality is too high for current 
 experiments.

We have summarized all our calculations and the significant experimental 
 results in Table~\ref{tab:res}.
The methods that we have used for calculating the upper bounds are described in 
 Appendix~\ref{app:table}.
It is important to remark that all mentioned experiments fail to satisfy 
 several of the conditions needed to extract loophole-free conclusions.
For example, all of them require the fair sampling assumption due to the low 
 detection efficiency.
Furthermore, in all of these experiments, locality is assumed rather than 
 enforced by spacelike separation.
Most critically, in all studied cases, the $n$-outcome measurements are 
 actually implemented using dichotomic measurements due to a limited number of 
 detectors.
But the existing experiments suggest that a loophole-free version of these 
 experiments is within current experimental reach and can demonstrate the 
 existence of genuinely nondichotomic and nontrichotomic quantum measurements.

At this point, the conclusion is that there is already evidence that there are 
 correlations in nature which cannot be explained by GPTs with dichotomic and 
 trichotomic quantum measurements.
However, more experimental effort is needed for a loophole-free confirmation of 
 this result, and even more theoretical and experimental effort is needed for 
 demonstrating correlations which cannot be explained by more general GPTs with 
 dichotomic measurements.

\textit{Probabilistic theories with $n$-chotomic measurements.}---%
Our main result, Theorem~\ref{thm}, establishes that nonsignaling $n$-chotomic 
 correlations $\boldsymbol P\in \mathcal P_n$ cannot explain all quantum 
 correlations.
In this section, we take the possibility seriously that QT does \emph{not} 
 account for correlations in nature and we argue how $n$-chotomic measurements 
 with fixed $n$ may constitute a plausible alternative to the construction used 
 in QT.

The first argument is the observation that, even in the everyday use of QT, we 
 find situations in which the set of actual measurements is only a subset of 
 the set of measurements possible \emph{a priori}.
One example is the superselection rules arising from the nonexistence of 
 certain ways of manipulating a system and the constraints on its time 
 evolution \cite{Wick:1952PR}.
Another example arises when quantum systems can only be manipulated locally.
There, the standard paradigm is the paradigm of local operations and classical 
 communication in which several separated parties have access to a shared 
 composite quantum system but there is no quantum interaction between the 
 parts.
Consequently, there are outcomes of two-outcome measurements that cannot 
 participate in certain measurements with more than two outcomes 
 \cite{Bennett:1999PRA, Kleinmann:2011PRA}.

The second argument why $n$-chotomic measurements may be a plausible 
 alternative to QT is its simplicity.
From the perspective of GPTs, the fact that a theory includes measurements 
 which cannot be created by choosing from two-outcome measurements is 
 surprising:
Any measurement with more than two outcomes can be coarse-grained to a 
 two-outcome measurement $(k,\text{not~}k)$, simply by only distinguishing 
 between the outcome labeled $k$ and any other outcome.
Now, consider the converse problem.
Suppose that we have the set of all two-outcome measurements of a GPT and we 
 want to construct the set of all measurements with any number of outcomes.
Then, the arguably simplest way to do it is as it is illustrated in 
 Fig.~\ref{fig:alice}, i.e., by selecting from two-outcome measurements.
The fact that this is not the case in QT tells us that QT is, in this sense, 
 very special.
Fortunately, Theorem~\ref{thm} shows that we can test whether nature is 
 special in this sense.

The third argument is that there is nothing \emph{a priori} problematic in a 
 dichotomic theory.
To illustrate this point, we develop a dichotomic theory based on QT.
For this purpose, it is enough to consider experiments consisting of two 
 stages, the preparation stage and the measurement stage.
In standard QT, a preparation is described by a density operator $\varrho$ and 
 a measurement by a family of positive semidefinite operators $(E_1, E_2, 
 \dotsc)$ summing to $\openone$, so that the probability to obtain outcome $k$ 
 is given by $\tr(E_k\varrho)$.

A straightforward example where two-outcome measurements are insufficient is a 
 measurement which can perfectly distinguish between more than two states so 
 that $\tr(\varrho_\ell E_k)= \delta_{\ell,k}$, where $\delta_{\ell,k}$ denotes 
 the Kronecker delta.
However, there is nothing particularly characteristic of QT in this example as 
 already in our everyday classical experience we can easily distinguish 
 different preparations---for example, the six distinct outcomes of a die.
In order to be able to separate this trivial example from the case we are 
 interested in, we consider a modification of QT.
Imagine that the state preparation does not only prepare the quantum state but, 
 in addition, transmits some information, e.g., an integer value $\alpha$.
In turn, the measurement apparatus is sensitive to $\alpha$ and can exhibit 
 different behavior depending on $\alpha$.
This means that $\alpha$ carries some classical information, e.g., which state 
 $\varrho_k$ was prepared or which side of the die is up, covering the 
 aforementioned situation, cf.\ also Fig.~\ref{fig:alice}.
In fact, this scenario is more realistic than it may seem.
For example, in a photon experiment, the halfwave plate used to prepare 
 different polarization states may introduce a slight shift in momentum, and it 
 may happen that the analyzing setup is sensible to this shift and gives a 
 different response depending on the momentum.

A general formalism to capture this situation is to modify the standard 
 formulation of QT by replacing the density operator $\varrho$ by positive 
 semidefinite operators $(\eta_1, \eta_2, \dotsc)\equiv \boldsymbol \eta$ 
 obeying $\sum_\alpha \tr(\eta_\alpha)= 1$ and to substitute each operator 
 $E_k$ by positive semidefinite operators $(D_{1,k},D_{2,k},\dotsc)\equiv 
 \boldsymbol D$ such that $\sum_k D_{\alpha,k}= \openone$ for each $\alpha$.
If there is no other sensitivity to $\alpha$, then outcome $k$ will have 
 probability $P(k)= \sum_\alpha \tr(\eta_\alpha D_{\alpha,k})$.
If we restrict the quantum part of the measurements to be trivial, i.e., all 
 $D_{\alpha,k}$ are either $\openone$ or $0$, then, effectively, we would have 
 a hidden variable model.
If, for each $\alpha$, at most two operators $D_{\alpha,k}$ are nonzero, then, 
 on a fundamental level, all measurements are dichotomic, and similarly for the 
 $n$-chotomic case.

Let us now use the above example to illustrate why at least bipartite 
 correlations are required to falsify these GPTs.
For a single party, we can always explain \emph{a posteriori} any experiment in 
 which the correlations of certain states $\boldsymbol \eta^{(\mu)}$ and 
 measurements $\boldsymbol D^{(\nu)}$ are considered.
Indeed, we may let $D^{(\nu)}_{\alpha,k}= p^{(\alpha,\nu)}_k$ and 
 $\eta^{(\mu)}_\alpha= \delta_{\alpha,\mu}$, where $p^{(\mu,\nu)}_k$ are 
 probability distributions that do not contradict the observations.
A way to inhibit such constructions is to move into a setup in which a system 
 is distributed between two parties and each of them performs local 
 measurements.
Then, instead of preparing states $\boldsymbol \eta^{(\mu)}$ and performing 
 measurements $\boldsymbol D^{(\nu)}$, both parties perform independent local 
 measurements $\boldsymbol D^{\prime(\mu)}$ and $\boldsymbol D^{(\nu)}$, 
 respectively, on a fixed bipartite state $\boldsymbol \eta$.
The resulting observations are then distributed according to the correlations
\begin{equation}
 P_{\mu,\nu}(k,\ell)= \sum_{\alpha',\alpha} \tr(\eta_{\alpha',\alpha} \,
 D_{\alpha',k}^{\prime(\mu)}\otimes D_{\alpha,\ell}^{(\nu)}).
\end{equation}
When all local measurements are at most $n$-chotomic, then, by construction, 
 these correlations are nonsignaling $n$-chotomic and are therefore subject to 
 Theorem~\ref{thm}.

\textit{Conclusions.}---%
Quantum theory (QT) is in agreement with all existing experimental evidence.
Therefore, when exploring alternative theories to QT, it is sensible to focus 
 on those giving similar predictions.
In this Letter we have studied a large class of such alternative theories.
We have considered a class of general probabilistic theories in which the set 
 of measurements is constructed by selecting from measurements with $n$ 
 outcomes.
For any $n$, these theories satisfy Bell-type inequalities which are violated 
 by QT.
Testing this prediction is a fundamental challenge for the future, as it would 
 demonstrate that correlations in nature are stronger than those allowed by 
 theories which, in other experiments, produce correlations exceeding those of 
 QT, e.g., as it is the case for Popescu–Rohrlich boxes.
However, this challenge is difficult and will require further efforts both in 
 theory and experiments.

Meanwhile, as an example of the kind of tools that will be needed, we have 
 considered theories with the same set of $n$-outcome measurements than QT for 
 a fixed $n$, but such that any measurement with more outcomes is constructed 
 by selecting measurements with only $n$ outcomes.
These theories share meany features with QT and can, e.g., explain the 
 violation of Bell inequalities.
However, we have shown that these alternative theories satisfy certain 
 Bell-type inequalities which are violated by QT.
The violations predicted by QT are very small and testing them requires 
 high-precision experiments.
We have identified previous experiments which, up to some assumptions, seem to 
 rule out these theories for $n=2$ and $n=3$.

\begin{acknowledgments}
We thank
Géza Giedke,
Gustavo Lima,
Géza Tóth, and
Tamás Vértesi
for discussions.
This work was supported by
the FQXi large grant project ``The Nature of Information in Sequential Quantum 
Measurements'',
project No.\ FIS2014-60843-P ``Advanced Quantum Information'' (MINECO, Spain) 
with FEDER funds,
the Project ``Photonic Quantum Information'' (Knut and Alice Wallenberg 
Foundation, Sweden),
the EU (ERC Starting Grant GEDENTQOPT), and the DFG (Forschungsstipendium KL 
2726/2–1).
\end{acknowledgments}

\appendix

\section{Proof of Theorem~\ref{thm}.}
\label{app:thm}
For the remaining step in the proof of Theorem~\ref{thm} we assume without loss 
 of generality that all measurement outcomes are labeled $k,\ell= 1,2,\dotsc$, 
 and we define $\mathcal P_{n,r}$ as the subset of $\mathcal P_n$ for which the 
 maximal index $k$ or $\ell$ is at most $r$.
We show that (a) $\inf[I'(\mathcal P_{n,r})]\ge 2^{1-r}$ for any $r$ and (b) 
 $I'(\mathcal P_{n,r'})= I'(\mathcal P_n)$ for some $r'$.
It follows that $\inf[I'(\mathcal P_n)]= \inf[I'(\mathcal P_{n,r'})]\ge 
 2^{1-r'}> 0$ holds, which is the statement needed in order to complete the 
 proof in the main text.

(a)
For arbitrary correlations $\boldsymbol P\in \mathcal P_{n,r}$ we denote by 
 $\boldsymbol P'\in \mathcal P_{n,r-1}$ the correlations where in $\boldsymbol 
 P$ the $r$th outcomes are merged with the first outcomes.
This implies
$
 P_{\mu,\nu}'(k\ge \ell)= P_{\mu,\nu}(k\ge \ell)+P_{\omitted,\nu}(r)
 -P_{\mu,\omitted}(r)+ P_{\mu,\nu}(r,1),
$
 and therefore,
$
 I'(\boldsymbol P')= I'(\boldsymbol P)- [
   P_{2,2}(r,1)+P_{1,2}(1,r) \quad\quad+P_{1,1}(r,1) -P_{2,1}(r,1)]
   \le I'(\boldsymbol P) +P_{2,1}(r,1) \le 2I'(\boldsymbol P).
$
By induction and due to $I'(\mathcal P_{n,1})= \set{1}$, we have $2^{1-r}\le 
 I'(\boldsymbol P)$.

(b)
We consider the set $\tilde{\mathcal P}_n$ of those correlations which can be 
 created from the correlations in $\mathcal P_{n,n}$ by applying all changes of 
 the labels of the outcomes $\lambda'_\mu\colon \set{1,\dotsc,n}\rightarrow 
 \naturals$, and similarly $\lambda_\nu$, via
$
 P_{\mu,\nu}(k,\ell)\mapsto P_{\mu,\nu}(\lambda'_\mu(k),
 \lambda_{\nu}(\ell)),
$
 while all other correlation terms are zero.
This does not yield more than $4n^2$ logical relations like 
 $\lambda_\mu(k)<\lambda_\nu(l)$ in $I'$ and hence, at most $2^{4n^2}$ 
 different labelings are needed to reach all logical relations.
Since this is a finite set, there is an integer $r'$ denoting the maximal 
 resulting index in the labelings, and therefore $I'(\mathcal 
 P_{n,r'})\supseteq
 I'(\tilde{\mathcal P}_n)$.
By definition, $\mathcal P_n$ is the convex hull of $\tilde{\mathcal P}_n$, so 
 that $I'(\tilde{\mathcal P}_n)= I'(\mathcal P_n)$ follows from the fact that 
 $I'$ is affine.
Therefore, $I'(\mathcal P_{n,r'})= I'(\mathcal P_n)$ holds due to $\mathcal 
P_{n,r'}\subseteq \mathcal P_n$.

\section{Quantum $n$-chotomic bounds in Table~\ref{tab:res}.}
\label{app:table}
The maximal quantum value is known for some inequalities or it can be 
 numerically approximated by a hierarchy of semidefinite programs suggested by 
 Navascués, Pironio, and Acín \cite{Navascues:2007PRL}.
For $n$-chotomic quantum measurements, one can proceed similarly, since it is 
 enough to maximize the value of the inequality, but with the additional 
 assumption that at most $n$ measurement outcomes are nontrivial.
By exploring all possible combinations with $n$ nontrivial outcomes and 
 calculating the maximal bound for each of these cases, we obtain the 
 $n$-chotomic bounds provided in Table~\ref{tab:res}.

We used the third level of the hierarchy for the values in the rows 
 ``{2}-outcome'' and ``{3}-outcome.''
Since this is an upper approximation on the true value, these values are at 
 most too pessimistic.
For the values in the row ``Quantum,'' the given values are for certain quantum 
 states and measurements.
This value is optimal for AN, and the value coincides with the bound from the 
 second level of the hierarchy for $I_3$ and $I_4$.
Only for VB, the third level of the hierarchy gives a slightly larger value, 
 $21.092> 21.090$.

\bibliography{the,xxx}

\end{document}